\documentclass[apj, numberedappendix]{emulateapj}

\usepackage{xspace}
\usepackage{color}
\usepackage{amsmath, bbm, ulem}

\shorttitle{UNITY}
\shortauthors{Rubin et al.}

\begin{document}

\newcommand{\nobs}{N^{\mathrm{obs}}\xspace}
\newcommand{\nmiss}{N^{\mathrm{miss}}\xspace}

\newcommand{\intr}{\mathrm{samp}}
\newcommand{\intrinsic}{sample\xspace}
\newcommand{\unmodeled}{unexplained\xspace}
\newcommand{\Unmodeled}{Unexplained\xspace}

\newcommand{\muBobsi}{\mu_{B i}^{\mathrm{obs}} \xspace}
\newcommand{\mBobsi}{m_{B i}^{\mathrm{obs}} \xspace}
\newcommand{\xoneobsi}{x_{1 i}^{\mathrm{obs}} \xspace}
\newcommand{\cobsi}{c_{i}^{\mathrm{obs}} \xspace}

\newcommand{\mBtruei}{m_{B i}^{\mathrm{true}} \xspace}
\newcommand{\xonetruei}{x_{1 i}^{\mathrm{true}} \xspace}
\newcommand{\ctruei}{c_{i}^{\mathrm{true}} \xspace}

\newcommand{\muBobs}{\mu_{B}^{\mathrm{obs}} \xspace}
\newcommand{\mBobs}{m_{B}^{\mathrm{obs}} \xspace}
\newcommand{\xoneobs}{x_{1}^{\mathrm{obs}} \xspace}
\newcommand{\cobs}{c^{\mathrm{obs}} \xspace}

\newcommand{\mBtrue}{m_{B}^{\mathrm{true}} \xspace}
\newcommand{\xonetrue}{x_{1}^{\mathrm{true}} \xspace}
\newcommand{\ctrue}{c^{\mathrm{true}} \xspace}

\newcommand{\mult}{\:}

\newcommand{\plusminus}[3]{$#1^{+#3}_{#2}$\xspace}

\newcommand{\capitalcoeff}{Standardization\xspace}
\newcommand{\coeff}{standardization\xspace}
\newcommand{\capitalcoeffs}{Standardizations\xspace}
\newcommand{\coeffs}{standardizations\xspace}

\newcommand{\nsimdata}{thirty\xspace}

\newcommand{\note}[1]{{\color{red}#1}}

\newcommand{\CZeroConstraint}{$c = -0.016 \pm 0.018$\xspace}
\newcommand{\dalphaconstraint}{\plusminus{-0.089}{-0.040}{0.035}\xspace}
\newcommand{\dbetaconstraint}{\plusminus{2.26}{-0.38}{0.43}}
\newcommand{\betablueconstraint}{\plusminus{0.56}{-0.34}{0.31}}
\newcommand{\betaredconstraint}{\plusminus{2.83}{-0.17}{0.18}}
\newcommand{\OmegamShiftMagLim}{only 0.006\xspace}
\newcommand{\Omegamnoskew}{only 0.001\xspace}
\newcommand{\outlierdistOmshift}{0.009\xspace}
\newcommand{\bayesianmodelsamedatashrink}{5\%\xspace}
\newcommand{\stepsevensmaller}{7\%\xspace}
\newcommand{\Omegamchangeafterunblinding}{0.009 ($0.2\sigma$)\xspace}

\title{UNITY: Confronting Supernova Cosmology's Statistical and Systematic Uncertainties in a Unified Bayesian Framework}

\author{
D. Rubin\altaffilmark{1, 2},
G. Aldering\altaffilmark{2},
K. Barbary\altaffilmark{2},
K. Boone\altaffilmark{2, 3},
G. Chappell\altaffilmark{1},
M. Currie\altaffilmark{1},
S. Deustua\altaffilmark{4},
P. Fagrelius\altaffilmark{2, 3},
A. Fruchter\altaffilmark{4},
B. Hayden\altaffilmark{2},
C. Lidman\altaffilmark{5},
J. Nordin\altaffilmark{6},
S. Perlmutter\altaffilmark{2, 3},
C. Saunders\altaffilmark{2, 3},
C. Sofiatti\altaffilmark{2, 3}
\\(The Supernova Cosmology Project)}

\altaffiltext{1}{Department of Physics, Florida State University, Tallahassee, FL, 32306}
\altaffiltext{2}{E.O. Lawrence Berkeley National Lab, 1 Cyclotron Rd., Berkeley, CA, 94720}
\altaffiltext{3}{Department of Physics, University of California Berkeley, Berkeley, CA 94720}
\altaffiltext{4}{Space Telescope Science Institute, 3700 San Martin Drive, Baltimore, MD 21218}
\altaffiltext{5}{Australian Astronomical Observatory, PO Box 296, Epping, NSW 1710, Australia}
\altaffiltext{6}{Institut f\"ur Physik, Newtonstr. 15, 12489 Berlin, Humboldt-Universit\"at zu Berlin}

\begin{abstract}
While recent supernova cosmology research has benefited from improved measurements, current analysis approaches are not statistically optimal and will prove insufficient for future surveys. This paper discusses the limitations of current supernova cosmological analyses in treating outliers, selection effects, shape- and color-\coeff relations, \unmodeled dispersion, and heterogeneous observations. We present a new Bayesian framework, called UNITY (Unified Nonlinear Inference for Type-Ia cosmologY), that incorporates significant improvements in our ability to confront these effects. We apply the framework to real supernova observations and demonstrate smaller statistical and systematic uncertainties. We verify earlier results that SNe Ia require nonlinear shape and color \coeffs, but we now include these nonlinear relations in a statistically well-justified way. This analysis was primarily performed blinded, in that the basic framework was first validated on simulated data before transitioning to real data. We also discuss possible extensions of the method.
\end{abstract}


\section{Introduction}

Recent supernova (SN) cosmological measurements have greatly reduced both the statistical and systematic uncertainty in our knowledge of the accelerated expansion of the universe \citep[the latest such efforts are presented in][]{suzuki12, betoule14, rest14}. Despite these improvements, the frameworks currently in use are not statistically optimal. As we build larger supernova samples, these frameworks will become increasingly inadequate.

This paper offers an improved technique for deriving constraints on cosmological parameters from SN measurements (peak magnitude, light-curve shape and color, host-galaxy mass). Although this paper uses SN cosmology as an example, researchers from other fields may find this type of framework useful. Our work is particularly relevant for researchers confronting partially known uncertainties, selection effects, correlated measurements, and outliers. The rest of this introduction summarizes the problems with existing methods and describes in general terms the basic requirements for greater accuracy. Section~\ref{sec:proposedframework} describes the framework in detail, and Section~\ref{sec:simulateddata} quantitatively examines the performance of the method on simulated data. In Section~\ref{sec:realdata}, we demonstrate the performance on real SN observations, then conclude in Section~\ref{sec:conclusions} with future directions.

\subsection{The Current Approach}\label{sec:currentapproach}

Current cosmological constraints are derived from time series of photometric measurements in multiple bands (light curves) and spectroscopy. Before obtaining SN distances, the light curves must be fit with a model such as SALT2 \citep{guy07, guy10, mosher14}. SALT2 models each photometric observation in the observer frame with a combination of a mean SN spectral energy distribution (SED) (scaled by a normalization parameter), the first component of the SED variation (scaled by a shape parameter $x_1$), and the mean color variation in magnitudes (which is scaled by a color parameter $c$). The template is also shifted in time to match the observations with a date-of-maximum parameter. (In this work, we restrict our attention to the SALT2 empirical model, which is the best validated and most widely used such model.) \citet{guy07, guy10} and \citet{mosher14} trained the SALT2 model in an initial (separate) step before the light-curve fitting, using a dataset with well-measured spectra and light curves that partially overlaps with the SN data used for cosmological distances.\footnote{We could, in principle, incorporate the training of SALT2 and the fitting of the light curves into our proposed model, but the necessary modeling of selection effects would be difficult.} After measuring a light-curve shape parameter ($\xoneobsi$), a light-curve color parameter ($\cobsi$), and a light-curve normalization ($\mBobsi$, the rest-frame $B$-band flux),\footnote{Of course, the real observables in astronomy are electrons per pixel for each image. We assume here that the photometry, calibration, and light-curve fitting yield an approximate multivariate Gaussian likelihood for each SN.} we construct a distance modulus estimate as

\begin{align} \label{eq:mBcorr}
\muBobsi & = \mBobsi + \alpha \mult \xoneobsi - \beta \mult \cobsi \notag \\
& + \delta \mult (M_{\star} > 10^{10} M_{\odot}) - M_B \;.
\end{align}

\noindent $\alpha$ and $\beta$ are the light-curve shape \coeff coefficient and color \coeff coefficient, respectively. They quantify the empirical relations that SNe with broader rest-frame optical light curves or bluer rest-frame colors are intrinsically more luminous \citep{phillips93, tripp98}. $\delta$ captures residual luminosity correlations with host-galaxy mass, discussed more in Section~\ref{sec:hostgalaxy}. $M_B$ is the absolute magnitude in the rest-frame $B$-band (for a given $H_0$). All of these coefficients are nuisance parameters for cosmological purposes.

We can then construct a $\chi^2$ to use for cosmological fitting (for illustrative purposes only, we assume here that all of the SN observations are uncorrelated):

\begin{equation} \label{eq:chi2fit}
\chi^2 = \sum_i \frac{ (\muBobsi(\alpha, \beta, \delta) - \mu(z_i, \mathrm{cosmo}))^2 } {\sigma^2_{\muBobs}(\alpha, \beta) + \sigma^{\intr^{\scriptstyle 2}}(\alpha, \beta) + \sigma^{\mathrm{ext} ^{\scriptstyle 2}}(z)} \;.
\end{equation}

\noindent $\sigma^2_{\muBobs}(\alpha, \beta)$ captures the measurement uncertainties from the light-curve fit. SALT2 handles $k$-corrections implicitly, as it is a rest-frame spectral model that fits photometry in the observer frame. Limitations of the SALT2 model are important to add, such as its inability to simultaneously model intrinsic color variation and extinction with one $c$ parameter (discussed more below). While SALT2 does include a simple estimate of the dispersion around its model, including this dispersion does not give a $\chi^2$ per degree of freedom of 1 when performing cosmological fits. $\sigma^{\intr^{\scriptstyle 2}}(\alpha, \beta)$ is a term that captures this sample-dependent \unmodeled variance in the SN distribution (sometimes called ``intrinsic dispersion''). Finally, $\sigma^{\mathrm{ext} ^{\scriptstyle 2}}(z)$ captures dispersion due to gravitational lensing and incoherent peculiar velocities. In current cosmological analyses \citep[those since][]{kowalski08}, an iteration is performed between estimating the \coeff coefficients, estimating $\sigma^{\intr^{\scriptstyle 2}}(\alpha, \beta)$, and rejecting outlying SNe.\footnote{There is a still some amount of confusion in the literature concerning Equation~\ref{eq:chi2fit}, and we address it now to contrast against the proposed framework for, e.g., fitting nonlinear $\{x_1,\ c\}$ standardizations. Finding the $\alpha$ and $\beta$ values that minimize this $\chi^2$ is not the same as minimizing the dispersion of the Hubble diagram. Nor does it eliminate the correlation between $\{x_1,\ c\}$ and Hubble diagram residuals. The residuals at the best-fit $\alpha$ and $\beta$ are expected to show a correlation with the observed $\{x_1,\ c\}$, as uncertainties will preferentially scatter the observed values of $\{x_1,\ c\}$ away from the distribution means and it is only the values without this scatter that should be uncorrelated with Hubble diagram residuals.}

\subsection{Limitations of Current Work}\label{sec:currentlimits}

There are several ways in which SN datasets are imperfect. Outliers, selection effects, nonlinear correlations, partially known uncertainties, and heterogeneity each complicate analyses. The proposed framework will address each of these complications, but first we discuss the limitations of current work.

Obtaining either spectra or very high-quality light curves are the only ways to ensure a transient source is a SN~Ia. Even at moderate redshifts, both of these techniques are observationally expensive, and non-Ia SNe will inevitably contaminate the sample. A similar issue arises if SNe are of type Ia, but are peculiar, or the redshift is incorrect. The analysis should thus accommodate some amount of non-Ia SNe, which have dissimilar colors, decline rates, and absolute magnitudes. The iterative outlier fit described above converges well for the sorts of contaminating distributions we expect when the samples are relatively pure \citep{kowalski08}. When the samples are large (several hundred), or impure ($\lesssim 85\%$)---conditions the field is beginning to face---the outliers can dominate the other sources of uncertainty in the fit. \cite{kunz07} presented a powerful Bayesian technique for simultaneously modeling the distributions of normal SNe and outliers, but does not confront many complexities of the data, including the luminosity \coeffs and selection effects.

Selection effects, the tendency for surveys to select against the faintest SNe, sculpt the observed distribution of SNe \citep[this is frequently referred to as Malmquist bias,][]{malmquist1922}. If not taken into account, this selection will bias both the cosmological parameter estimation and the \coeff coefficients. There are SN-to-SN variations in the selection efficiency, even within the same survey and for the same redshift (e.g., due to seeing, host-galaxy contamination, or moonlight). These are deterministic but difficult-to-model sources of variation. Noise also plays a role, as a SN right at the detection threshold may stochastically scatter above or below it. Thus, the detection efficiency transitions to zero smoothly as a function of apparent brightness. In most SN cosmological analyses (based on Equation~\ref{eq:chi2fit}), the treatment of selection effects is performed outside the statistical framework, in the form of survey simulations followed by an ad-hoc redshift-dependent adjustment of the $\mBobsi$ to approximately cancel the estimated bias \citep{kessler09, conley11}. Bayesian analyses face a related challenge: selection effects also influence the distribution of light-curve width and color with redshift, which can amplify selection effects if not modeled \citep{woodvasey07}. However, these frequentist and Bayesian analyses both require knowledge of the true population distribution (as a function of redshift) to get accurate results.

The $x_1$ and $c$ \coeffs in Equation~\ref{eq:mBcorr} are linear. However, nonlinear decline-rate and color \coeffs are statistically justified \citep{amanullah10, scolnic14b, scolnic14a}.\footnote{Others have seen a trend towards lower color-\coeff coefficients for even redder SNe, $c \gtrsim 0.5$ \citep{mandel11, burns14}, but virtually all the SNe used for cosmology are bluer than this.} Our Union2 result \citep{amanullah10} was derived from subdividing the sample by the best-fit latent variables (see Section~\ref{sec:latentvariables}). The subdivisions showed very similar cosmological results, but subdivision tests are statistically weak (statistical uncertainties on the difference are $\sim 2$ times larger than the uncertainty on the whole sample). Including these nonlinear \coeffs in the fit would be preferable for evaluating their impact and would remove any bias created by the assumption of linearity. Unfortunately, neither the \citet{amanullah10}, nor the \citet{scolnic14a} frameworks were able to incorporate these nonlinear \coeffs in a fully self-consistent way.

Type Ia Hubble diagrams show more dispersion in distance than can be explained with measurement uncertainty alone. As noted above, the framework must include a model of the \unmodeled dispersion, and should not assume that all of this dispersion is in the independent variable (magnitude). In other words,  $\sigma^{\intr^{\scriptstyle 2}}$ is really $\sigma^{\intr^{\scriptstyle 2}}(\alpha, \beta)$.\footnote{The SALT2 light-curve fitter already incorporates a significant amount of model uncertainty in both the color and magnitude measurements.} SN variation (beyond light-curve width and color), is presently only crudely modeled by current light-curve fitters and especially affects the measured color \citep{wang09, chotard11, foley11a, foley11b, foley12b, saunders15}, resulting in color measurements that mix extinction and SN differences in only partially understood ways. When using Equation~\ref{eq:chi2fit}, there is no way to fit for the \unmodeled dispersion simultaneously with the other parameters. Improvement is needed here, as the \unmodeled dispersion interacts with the modeling of selection effects, the outlier rejection, and fitting $\alpha$ and $\beta$. Currently, we are constrained to use ad-hoc methods, such as performing many fits with a randomized \unmodeled-dispersion matrix, and computing the distributions of cosmological parameters fit to fit \citep{marriner11}. The impact of these variations is currently subdominant to the total uncertainty \citep{kessler13, mosher14}, but this may not be true in future samples and a precise way to evaluate the cosmological impact is desirable. \citet{march11} took a step in the right direction by using a Bayesian hierarchical model to simultaneously fit for cosmological parameters and \unmodeled dispersion, but their model cannot accommodate calibration uncertainties, outliers, nonlinear \coeffs, or selection effects.

Future SN cosmology analyses will confront additional difficulties, such as heterogeneity. Even very homogeneous SN imaging surveys \citep[such as the Dark Energy Survey, ][]{flaugher05, bernstein12} will gain heterogeneity as expensive observational resources, such as near-IR measurements or high-quality spectroscopy, will only be available for a subset of objects. A frequentist, ``object-by-object'' uncertainty analysis such as Equation~\ref{eq:chi2fit} cannot make efficient use of this information.\footnote{The Union2.1 analysis \citep{suzuki12} already uses a simple Bayesian model for the host-galaxy mass \coeff, where samples which lack host-mass measurements are given priors derived from other similar samples, but the remainder of that analysis uses a frequentist framework.}

\subsection{Desired Properties for a New Approach} \label{sec:desiredprops}

In light of these problems, this paper offers an improved technique incorporating a more sophisticated, Bayesian model of the data. Properly making use of heterogeneous information (like measurements that are only available for a subset of objects) requires a model of the SN population in which the parameters of the distribution are treated as unknowns. In addition, the model of the \unmodeled dispersion should allow for uncertainty in both size and functional form. We discuss the details of our procedure for parameterizing these possibilities in Section~\ref{sec:intdisp}. Only a Bayesian framework can accommodate more exotic possibilities, such as very large numbers of nuisance parameters (each possibly non-zero and having initially unknown size), as it can make use of a hierarchical prior around zero.\footnote{In that kind of Bayesian hierarchal model, the width of the prior around zero is incorporated as a hyperparameter in the fit. The prior will smoothly weaken as the statistical evidence builds that some of these nuisance parameters are non-zero. If there is no evidence that any of these parameters are non-zero, the prior will remain clustered around zero.} One could imagine many possibilities for these parameters (e.g., color-\coeff coefficient, $\beta$, as a function of host-galaxy-inclination angle); we do not pursue these in this work, but we do note that they could be built into the general framework.

However, not all of the improvements we propose require Bayesian statistics; some are due to the improvements to the data likelihood (and could thus fit into a frequentist framework). First, we introduce a modified approach to fitting for residual correlations with host-galaxy mass, described in Section~\ref{sec:hostgalaxy}. Second, like \citet{kunz07}, we handle outliers with a mixture model, described in Section~\ref{sec:nonIa}, which offers improved robustness. Third, to account for selection effects, our proposed framework uses a probabilistic model of selection as described in Section~\ref{sec:selection}, modifying the classical likelihood to include selection directly. This cleanly estimates and marginalizes the hyperparameters of the true population distributions simultaneously with other parameters, propagating all uncertainties. Fourth, in Section~\ref{sec:indpriors}, we discuss our approach to fitting for the changing SN independent variable distribution with redshift (population drift) simultaneously with other parameters. Finally, our new framework can accommodate nonlinear \coeffs (Section~\ref{sec:brokenlinear}). A frequentist framework can, in principle, include nonlinear \coeffs, but it is computationally challenging.\footnote{Note that the likelihoods for the latent variables can easily become multi-modal for nonlinear \coeffs, as a curved line may approach a datapoint more than once. For this reason, it is more practical to approach the problem with a technique like MCMC than a derivative-based minimizer. The lead author has had success in a frequentist framework only by testing each local $\chi^2$ minimum for each latent variable and recording the best one.}

Both the old and new techniques can handle correlations due to systematic uncertainties (such as correlated peculiar velocities and photometric calibration uncertainties), assuming the sizes of these uncertainties are known.\footnote{There is a benefit to a Bayesian approach if the systematic uncertainties have unknown sizes, and these sizes must be marginalized over.} The modifications to the old framework to include correlations are detailed in \citet{kowalski08, amanullah10, guy10, conley11}. We implement these correlations with nuisance parameters, as discussed in Section~\ref{sec:systerrs}. The model must also allow for population drift (whether due to selection effects, or changes in the SN population with redshift), or risk a significant bias on cosmological parameters. Population drift can also be handled well with either framework, as Equation~\ref{eq:chi2fit} handles each object independently (if the measurements are uncorrelated object to object) and our new framework solves for population drift simultaneously with other parameters (Section~\ref{sec:indpriors}).

There are two disadvantages to our approach. First, the more sophisticated model requires more CPU time (now measured in many hours, rather than minutes). Second, it is not necessarily possible to generate a unique distance modulus for each SN as in Equation~\ref{eq:mBcorr}. The true-independent-variable priors (Section~\ref{sec:indpriors}) must be treated in bins smaller than the other parameters of interest (e.g., bins of $\lesssim 0.1$ in redshift for cosmological parameters). However, no one set of assumptions will exactly cover all use cases (e.g., tests for isotropy). Approximate sets of SN distances are possible, but we leave this for future work.

\section{Proposed Method} \label{sec:proposedframework}

\begin{figure*}[Ht]
\begin{center}
\includegraphics[width = 0.6 \textwidth]{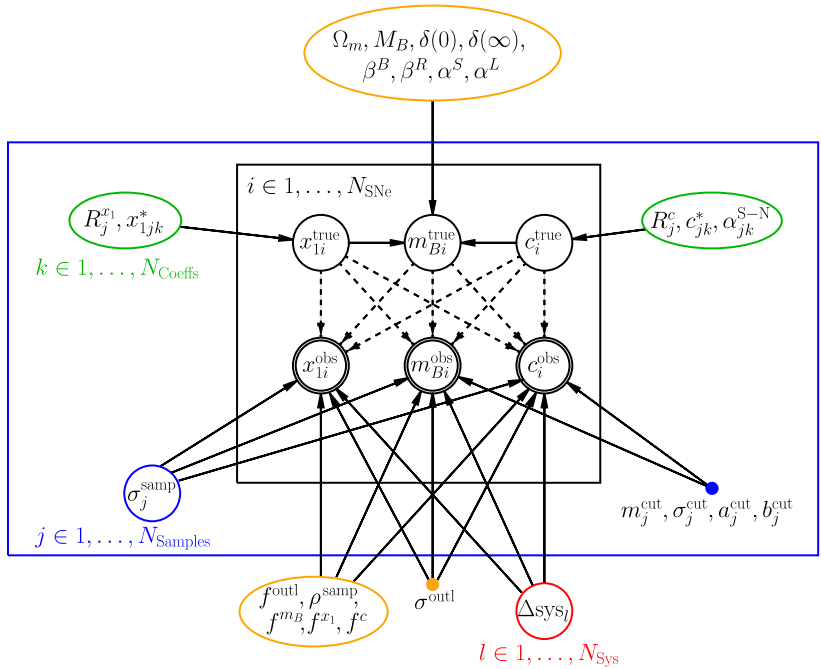}
\end{center}

\caption{Probabilistic Graphical Model of our framework showing the causal links. An edge from one node (e.g., $\Omega_m$) to another (e.g., $\mBtruei$) means that the latter is conditional upon the former (e.g., $\mBtruei$ is conditional on $\Omega_m$). The enclosed nodes represent variables that are sampled in the MCMC. Global parameters are in orange nodes (single parameters) and red nodes (the set of systematic uncertainty parameters). Green nodes enclose the hyperparameters (parameters of a prior distribution) of the latent-variable priors, and the singly outlined black nodes show those latent variables. Blue nodes show sample-dependent quantities. Finally, the outlined nodes show the observed light-curve fits. (Each of $\{\mBobsi,\ \xoneobsi,\ \cobsi\}$ depends on $\{\mBtruei,\ \xonetruei,\ \ctruei\}$ as the light-curve fit and \unmodeled dispersion have correlated uncertainties.) $i$ ranges over each SN, $j$ ranges over each SN sample, $k$ ranges over the coefficients in redshift within a sample, and $l$ ranges over each systematic uncertainty (e.g., calibration). Note that the $\mBtruei$ are completely determined by other parameters and are not true fit parameters. We fix the selection effect parameters, $m_j^{\mathrm{cut}}$, $\sigma_j^{\mathrm{cut}}$, $a_j^{\mathrm{cut}}$, $b_j^{\mathrm{cut}}$, and the outlier distribution width $\sigma^{\mathrm{outl}}$, so these are represented filled nodes.}
\label{fig:DAG}
\end{figure*}

Our proposed framework surpasses previous analysis efforts by bringing together components that simultaneously address each of the limitations discussed above. We call our framework UNITY (Unified Nonlinear Inference for Type-Ia cosmologY). The required model needs to contain thousands of nonlinear fit parameters (as motivated in the following sections), which poses a problem for many statistical techniques. For example, there is no practical way to analytically marginalize over every parameter except the parameter(s) of interest. We thus must draw random samples from the posterior distribution (Monte Carlo sampling), and use those samples to estimate the posteriors. A natural tool for this sampling is Hamiltonian Monte Carlo (HMC). HMC samples efficiently, with short correlation lengths sample to sample, even for large numbers of fit parameters. To this end, we use Stan \citep{Hoffman-Gelman:2011, hoffman-gelman:2013, betancourt:2013} through the PyStan interface \citep{pystan-software:2014}. Stan automatically chooses a mass matrix, and speeds up sampling efficiency even further by using a variant of HMC sampling called the ``no-U-turn'' sampler. Stan also incorporates automatic, analytic differentiation for computing the gradient of the log-likelihood, making the implementation of the model simple and readable. We show a Probabilistic Graphical Model of our framework in Figure~\ref{fig:DAG}, and show a table of parameters in Table~\ref{tab:paramlist}.

\begin{deluxetable*}{llr}[H]
\tabletypesize{\scriptsize}
\tablecaption{Parameters in our model.}
\tablewidth{0pt}
\tablehead{
\colhead{Parameters} & \colhead{Symbols} & \colhead{Section}
}
\startdata
Absolute Magnitude for $h=0.7$ & $M_B$ & \ref{sec:currentapproach} \\[0.02in]
Cosmological Model (Flat $\Lambda$CDM) & $\Omega_m$ & \ref{sec:currentapproach} \\[0.02in]
Latent Variables & $\xonetruei$, $\ctruei$ & \ref{sec:latentvariables} \\[0.02 in]
Host-Mass \capitalcoeff Coefficients & $\delta(0)$, $\delta(\infty)$ & \ref{sec:hostgalaxy}\\[0.02 in]
Outlier Distribution & $f^{\mathrm{outl}}$, $\sigma^{\mathrm{outl}}$ & \ref{sec:nonIa}\\[0.02 in]
Sample Limiting Magnitudes (fixed) & $m_j^{\mathrm{cut}}$, $\sigma_j^{\mathrm{cut}}$, $a_j^{\mathrm{cut}}$, $b_j^{\mathrm{cut}}$ & \ref{sec:selection} \\[0.02 in]
Latent-Variable Hyperparameters & $R_j^{x_1}$, $x_{1jk}^*$, $R_j^c$, $c_{jk}^*$, $\alpha_{jk}^{\mathrm{S-N}}$& \ref{sec:indpriors} \\[0.02 in]
Light-Curve Color and Shape \capitalcoeff Coefficients & $\beta^B$, $\beta^R$, $\alpha^S$, $\alpha^L$ & \ref{sec:brokenlinear} \\[0.02 in]
`Sample'' (\Unmodeled) Dispersion & $\sigma_j^{\intr}$, $f^{m_B}$, $f^{x_1}$, $f^c$, $\rho^{\intr}$ & \ref{sec:intdisp} \\[0.02 in]
Systematic Uncertainties & $\Delta\mathrm{sys}_l$ & \ref{sec:systerrs}
\enddata 
\tablecomments{The final column displays section references for the parameters in our model.}
\label{tab:paramlist}
\end{deluxetable*}

\subsection{Parameterization of the True Position on the \capitalcoeff Relation}\label{sec:latentvariables}

\newcommand{\yobs}{\ensuremath{y^{\mathrm{obs}}}\xspace}
\newcommand{\xobs}{\ensuremath{x^{\mathrm{obs}}}\xspace}
\newcommand{\ytrue}{\ensuremath{y^{\mathrm{true}}}\xspace}
\newcommand{\xtrue}{\ensuremath{x^{\mathrm{true}}}\xspace}

\citet{gull89} offers an excellent discussion of linear regression with error bars in both dependent and independent variables. We briefly summarize here.
Consider the case of fitting a straight line (slope $A$, intercept $B$) in two dimensions ($y$, the dependent variable vs $x$, the independent variable) with uncertainties in both $x$ and $y$ ($\sigma_x$ and $\sigma_y$, assumed uncorrelated and Gaussian for simplicity). The generative model for an observed $y$ (``\yobs'') is
\begin{equation}
\yobs \sim \mathcal{N}(\ytrue,\ \sigma_y^2)
\end{equation}
and similarly for $\xobs$
\begin{equation}
\xobs \sim \mathcal{N}(\xtrue,\ \sigma_x^2)\;, \label{eq:xtrueexample}
\end{equation}
where \xtrue and \ytrue are the value for the measurement if no uncertainty is present, and $\ytrue = A\mult \xtrue + B$. When there is no significant uncertainty in $x$, Equation~\ref{eq:xtrueexample} is unnecessary, and we have a simple least-squares regression. When uncertainty is present in both variables, we can substitute for \ytrue, but \xtrue remains. For a fit with two independent variables, there are two latent variables representing the true position on the line (now a plane). The same logic holds for more than one observation, in which case there are (Number~of~Observations)$\times$(Number~of~Independent~Variables) latent variables. For the standard $x_1$ and $c$ \coeffs, this results in $2N_{\mathrm{SNe}}$ additional parameters.\footnote{If the redshifts of the SNe are also unknown, the analysis will instead face $3N_{\mathrm{SNe}}$ parameters. Because of the potential of misassociation, this situation might be faced by surveys measuring many host-galaxy redshifts after the SNe have faded, such as the Dark Energy Survey.}

In a frequentist analysis, if measurement uncertainties are believed to be Gaussian and the $x_1$ and $c$ \coeffs are linear, the likelihood can be analytically maximized for each of these parameters \citep{kowalski08}. This technique enables the proper handling of these nuisance parameters without explicitly including them in the fit. Similarly, in a Bayesian analysis, if the measurement uncertainties are Gaussian (or a Gaussian mixture), the \coeffs are linear, and the priors on these parameters are flat or Gaussian (or a Gaussian mixture), these nuisance parameters can be analytically marginalized \citep{gull89}, providing a similar computational efficiency boost \citep[as was done in][]{march11}. In this work, we violate these assumptions; therefore, we must keep these $2N_{\mathrm{SNe}}$ nuisance parameters in the fit ($\xonetruei$ and $\ctruei$).

\subsection{Host-Galaxy Environment Standardization Relation} \label{sec:hostgalaxy}

\cite{kelly10} first found evidence that Hubble residuals in current light-curve fitters are correlated with host-galaxy environment. This finding later reached high statistical significance in larger samples \citep{sullivan10, lampeitl10, childress13}. The current method for including this effect in cosmological analyses fits two separate absolute magnitudes for low-mass-hosted ($M_* < 10^{10} M_{\odot}$) and high-mass-hosted galaxies \citep{sullivan10, suzuki12}. \citet{rigault13} found evidence that most of this effect is due to the age of the progenitor system \citep[see also indirect evidence in ][]{childress13}. This effect, confirmed independently by \citet{rigault15}, implies that SNe hosted by high-stellar-mass galaxies may become more like low-stellar-mass-hosted SNe at higher redshift (when progenitors were young in all galaxies). However, newer versions of SALT2 combined with different sample selections may not show as strong an age effect \citep{rigault15, jones15}.

We therefore do not assume a constant mass-\coeff ($\delta$); instead we use a modified version of the model in \citet{rigault13} \citep[similar to the model of ][]{childress14}

\begin{align}
M_B^{\mathrm{low}} - M_B^{\mathrm{high}} & \equiv \delta(z) \notag \\
& = \delta(0) \left[ \frac{1.9 \left(1 - \frac{\delta(\infty)}{\delta(0)} \right)}{0.9 + 10^{0.95 z}}  + \frac{\delta(\infty)}{\delta(0)} \right] \;. \label{eq:deltaofz}
\end{align}

Those authors' proposed host-mass-\coeff evolution predicts the mass-\coeff coefficient will approach zero at high-redshift; we instead assume it smoothly approaches a possibly non-zero quantity, $\delta(\infty)$. We take a flat prior on $\delta(\infty)/\delta(0)$ from 0 to 1, allowing the mass \coeff to be constant or declining with redshift, and spanning all of the claims in the literature.

\subsection{Non-Ia or Other Outlier Contamination} \label{sec:nonIa}

For non-Ia contamination, we use the mixture-model framework of \citet{kunz07}. This framework models the observed distribution around the modeled mean as a sum of Gaussians, where at least one Gaussian (the normal Ia distribution) is tightly clustered, and an outlying distribution is comparatively dispersed. Although the assumption of a Gaussian contaminating distribution is a strong one, it makes little difference in practice. As the outlying distribution is broader than the inlying distribution, any outlying point will be treated as an outlier. For the relatively pure spectroscopically confirmed SN datasets in use today, modeling the outlying distribution accurately has little impact on the rest of the parameters in the model. Because of this, and as a test of the framework, we perform our fits assuming an outlier distribution that is a unit multinormal ($\sigma^{\mathrm{outl}} = 1$) in $\{m_B,\ x_1,\ c\}$ (centered on the Ia distribution for that redshift), which is different from the simulated data we generate to test the framework (Section~\ref{sec:simulateddata}).
 
The relative normalizations of the core and outlying distributions can be chosen object by object (from spectroscopic or other classification evidence), or set to be the same for every object. We fit for the fraction of outliers assuming it is the same for all objects ($f^{\mathrm{outl}}$), and place a broad log-normal prior on this quantity of $-3 \pm 0.5$ (an outlier fraction of $\exp(-3) = 0.05$ plus or minus 50\%). These assumptions work well with Union2.1, as discussed in Section~\ref{sec:modelwithoutl}, but of course they can be adjusted for other datasets.

If the luminosity distribution of SNe Ia turns out to be significantly non-Gaussian \citep[for example, the bimodal model of ][]{rigault13}, additional Gaussian components can be added (with redshift-dependent normalizations) to give smaller uncertainties and capture possible population drift. We leave this and more complex models of the non-Ia distribution for future work, but these extensions fit easily into this framework.

\subsection{Selection Effects} \label{sec:selection}

We present the details of our selection model in Appendix~\ref{sec:detailsselectioneffects}, but outline the important points here. The standard method for incorporating a selection cut is to truncate the data likelihood at the cut and divide by the selection efficiency (e.g., \citealt{gelman2013bayesian}, see also \citealt{kelly07} for a discussion of selection effects and non-detections in the context of linear regression). In SN cosmology, the truncation is not sharp, but is instead probabilistic, as discussed in Section~\ref{sec:currentlimits}. We assume that the observation likelihood is truncated by an error function. Far from the selection limit, a SN is found or missed with probability one or zero; for SNe near the selection limit, the probability transitions smoothly. An error function reasonably matches the efficiency curves of e.g., \citet{dilday08, barbary12, perrett12, graur14, rodney14}.\footnote{An error-function truncation rapidly approaches 100\% efficiency on the bright side of the cut. Real surveys are not as ideal, and will asymptote short of 100\%. As this asymptote has little brightness dependence, it does not impact the selected population.}

Surveys also do not select only on one measured variable. We assume our cut is a plane in three-dimensional space, spanning the dependent variable and both independent variables. In our example, this is magnitude (SALT2 $m_B$), shape (SALT2 $x_1$), and color (SALT2 $c$); SNe with $m_B + a^{\mathrm{cut}} x_1 + b^{\mathrm{cut}} c > m^{\mathrm{cut}}$ are less than 50\% likely to be found (and more than 50\% likely to be found if $< m^{\mathrm{cut}}$). The width of the cut is $\sigma^{\mathrm{cut}}$; SNe observed at $m^{\mathrm{cut}} \pm \sigma^{\mathrm{cut}}$ are considered 16\%(+) or 84\%($-$) likely to be found. $m_B$ and $c$ are the primary variables responsible for selection effects in SN searches. For example, SNe found in the rest-frame $B$-band have a limit in $m_B$ ($ b^{\mathrm{cut}}=0$), while SNe selected in the rest-frame $V$-band have a limit in $m_B - c \approx m_V$ or $ b^{\mathrm{cut}}=-1$. We note that for selection in just $m_B$, bluer (more-negative $c$) and slower-declining SNe (larger $x_1$) will be selected, as these correlate with brighter $m_B$. That is, the only effect that a magnitude-based $\{m_B,\ c\}$ selection ignores is that slower-declining SNe are more likely to be found, irrespective of the maximum brightness, as they stay above the detection threshold longer. A simple simulation shows that this effect is very small compared to selection on $\{m_B,\ c\}$, even for cadences as large as ten rest-frame days. Another bias related to selection effects is the bias due to larger uncertainties on fainter SNe \citep[e.g.,][]{kowalski08}. Simple simulations show that our Bayesian framework has much less susceptibility to this bias, and the uncertainty bias is much smaller than the one due to missing faint SNe altogether.

\subsection{Priors on True Values of Independent Variables} \label{sec:indpriors}

As this is a Bayesian framework, we must select priors on the true $x_1$ and $c$ latent parameters (see \citealt{gull89} for a discussion of Gaussian priors, and \citealt{kelly07} for a Gaussian-mixture prior). These priors must be chosen very carefully. If the prior mean is wrong, then every distance will also be incorrect in a correlated direction. The variance of the prior has an impact as well. If the prior variance is larger than the population variance, then the true latent parameters will be scattered about the mean more than they should, and the slope of the line will be biased towards zero. The converse will bias the slope of the lines away from zero. The mean and variance of the prior are the most important parameters to estimate accurately, thus Gaussians are normally adequate. In SN cosmology, these priors must also be redshift-dependent as the SN population can drift with redshift.

The optimal way to ensure the proper size and redshift-dependence of the priors is to fit for the prior parameters (the ``hyperparameters'') simultaneously with every other parameter. We selected skew normal distributions for the $c$ priors (allowing the distribution to be skewed), and Gaussians for the $x_1$ priors. The prior must be able to vary in redshift more rapidly than the cosmological fit in order to not introduce a bias. What we propose here more than meets this mild requirement,\footnote{In contrast, the constant-in-redshift priors of \citet{march11} do not meet this requirement.} but there is no harm in allowing the hyperparameters to mimic more closely the redshift dependence of the $x_1$ and $c$ distributions. For each sample, we fit for the mean of the distributions as a function of redshift with a linear spline. We use up to four spline nodes (the $x_{1jk}^*$ and $c_{jk}^*$, for $x_1$ and $c$, respectively), equally spaced in redshift over the range of a sample, with linear interpolation between these nodes.\footnote{Specifically, we use two nodes (a single line segment) for datasets with fewer than 30 SNe; we use three for 30-39 and four when the number of SNe is at least 40. This improves the robustness of the determination of the independent variable standard deviations.} We take non-informative flat priors on the means of the distributions and on the log of the standard deviations (the standard deviations for each sample $j$ are $R_j^{x_1}$ and $R_j^c$).\footnote{For a skew normal distribution, the mean and standard deviation are not the same quantities as $\mu$ and $\sigma$.} For the shape parameter $\alpha^{\mathrm{S-N}}$ of the skew normals, we take a flat prior on $\delta^{\mathrm{S-N}} = \alpha^{\mathrm{S-N}}/\sqrt{1 + (\alpha^{\mathrm{S-N}})^2}$, which is also allowed to vary in redshift in the same way as the distribution mean. This prior forces the skew normal to approach a Gaussian for samples with few objects. (The superscript ``S$-$N'' here is used to distinguish these skew-normal variables from the SN \coeff coefficients.)

For simplicity, we assume that the true $x_1$ and $c$ distributions are uncorrelated (the observed distributions of $x_1$ and $c$ show little correlation). We do note that ignoring correlations will bias the fit if any significant ($|\rho| \gtrsim 0.2$) correlations are present. 

\subsection{Nonlinear Independent Variable \coeffs} \label{sec:brokenlinear}

With the true values of the independent variables explicit in the model, it becomes trivial to have nonlinear \coeffs. For this work, we suggest a broken-linear relationship, allowing red/blue and small-/large-$x_1$ SNe to have different size \coeff coefficients ($\beta^R$/$\beta^B$ and $\alpha^S$/$\alpha^L$, respectively). We take a flat prior on the angle of each line segment, but transform to the average slope and the difference in slopes for display purposes: $\alpha \equiv (\alpha^L + \alpha^S)/2$, $d\alpha \equiv \alpha^L - \alpha^S$, $\beta \equiv (\beta^R + \beta^B)/2$, and $d\beta \equiv \beta^R - \beta^B$. Although the $x_1$ and $c$ values of the break could be fit parameters, we do not do this for our primary fits. For the moment, we split the sample at $x_1$ and $c$ of 0.

\subsection{\Unmodeled Dispersion} \label{sec:intdisp}

\newcommand{\sigint}{\sigma_j^{\intr}\xspace}
As the \unmodeled dispersion is parameterized in the model, it can be marginalized. We do not know what functional form to assume, so we use a flexible parameterization. Each SN sample is allowed its own \unmodeled dispersion, allowing poorer-quality samples to be naturally deweighted. We must also distribute the \unmodeled dispersion over $m_B$, $x_1$, and $c$, while accounting for possible correlations.

First, we split the variance of the \unmodeled dispersion into $f^{m_B}$, $f^{x_1}$, and $f^{c}$ (the fraction of the \unmodeled variance in $m_B$, $x_1$, and $c$, respectively), which are constrained to sum to one. Then, we scale each of these by 1, $0.13^{-2}$, and $(-3.0)^{-2}$. The values $0.13$ and $-3.0$ approximately scale out $\alpha$ and $-\beta$, respectively, where the negative sign for $\beta$ corresponds to the sign convention in Equation~\ref{eq:mBcorr}. (Note that using $\alpha$ and $-\beta$ directly would cancel the $\alpha$ or $\beta$ dependence when computing a marginalized distance uncertainty for each SN, so this would be inappropriate.) We also scale the variance by the \unmodeled dispersion for each sample, $\sigma_j^{2 \intr}$. Finally, we form a covariance matrix out of this \{$m_B$,~$x_1$,~$c$\} \unmodeled variance, allowing the off-diagonals to be scaled by parameters $\rho^{\intr}$, as follows

\begin{equation} \label{eq:fracvec}
\left(
\begin{array}{ccc}
f^{m_B} & \rho^{\intr}_{12} \frac{\sqrt{f^{m_B} f^{x_1}}} {0.13} & \rho^{\intr}_{13} \frac{\sqrt{f^{m_B} f^{c}}} {-3.0} \\
\rho^{\intr}_{12} \frac{\sqrt{f^{m_B} f^{x_1}}}{0.13}  & \frac{f^{x_1}}{0.13^2}& \rho^{\intr}_{23} \frac{\sqrt{f^{x_1} f^{c}}}{(-3.0)(0.13)} \\
\rho^{\intr}_{13} \frac{\sqrt{f^{m_B} f^{c}}} {-3.0} & \rho^{\intr}_{23} \frac{\sqrt{f^{x_1} f^{c}}}{(-3.0)(0.13)}  & \frac{f^{c}}{(-3.0)^2} 
\end{array}\right) \sigma^{\intr^{\scriptstyle 2}}_j 
\end{equation}

We take a non-informative ``LKJ'' prior on the correlation distribution \citep{lewandowski09}, with $\eta = 1$, as well as a flat prior on $\log{\sigint}$.

\subsection{Treatment of Correlated SN Observations} \label{sec:systerrs}

Many effects result in correlated measurements in SN cosmology. The most notable such effect results from common calibration paths: SNe from a given dataset share systematics such as telescope bandpass and photometric calibration uncertainties. Other effects include correlated uncertainties in Milky Way extinction maps, and correlated peculiar velocities. In standard analyses, these effects are propagated into a covariance matrix \citep{amanullah10, guy10, conley11}.

In order to speed up the Monte-Carlo sampling, we leave each correlating factor explicit as a parameter \citep[similar to][]{kowalski08}, $\Delta{\mathrm{sys}}_l$ (where $l$ ranges over all systematic uncertainties), while leaving the data uncorrelated SN to SN. These two approaches coincide exactly for a linear model with Gaussian uncertainties \citep{amanullah10}. The $\Delta{\mathrm{sys}}_l$ parameters capture the deviations of a measured quantity, like a zeropoint or a filter bandpass, from the estimated value. For each quantity, we numerically compute the derivative of the light-curve fit ${\mBobsi,\ \xoneobsi,\ \cobsi}$ with respect to that quantity, giving each $\partial \mBobsi/ \partial \Delta{\mathrm{sys}}_l$, $\partial \xoneobsi/ \partial \Delta{\mathrm{sys}}_l$, and $\partial \cobsi/ \partial \Delta{\mathrm{sys}}_l$. This lets us marginalize out $\Delta{\mathrm{sys}}_l$, with a Gaussian prior around zero set by the estimated size of each systematic uncertainty.

\section{Simulated Data Testing}\label{sec:simulateddata}

Our analysis framework must pass through careful testing using simulated data before it can be applied to real data. To this end, we generate \nsimdata simulated datasets that incorporate many characteristics of real data. As our analysis takes the SALT2 light-curve fits as inputs, this is the level at which we generate simulated data. 

\begin{itemize}
\item We generate four simulated datasets spanning the redshift ranges 0.02--0.05, 0.05--0.4, 0.2--1.0, and 0.7--1.4.
\item Each simulated dataset has 250 SNe, except the highest-redshift, which has 50.
\item We generate the $x_1$ population from a unit normal distribution, centered on zero.
\item We draw the population $c$ values from the sum of a Gaussian distribution of width 0.1 magnitudes, and an exponential with rate $1/(0.1 \mathrm{\; magnitudes})$. We center the distribution on zero.
\item We assume that the \unmodeled dispersion covariance matrix is correct in SALT2, and that only dispersion in $m_B$ (gray dispersion) remains. The statistical model does not have access to this information, and fits for the full unknown matrix, overestimating the uncertainties on $x_1$ and $c$, and thus slightly biasing $\alpha$ and $\beta$ away from zero (see Section~\ref{sec:indpriors}). (This is not a unique problem for our framework; the old technique would have the same bias.)
\item We assume that the uncertainties on $m_B$, $x_1$, and $c$ are 0.05, 0.5, and 0.05, and are uncorrelated. In addition, we take 0.1 magnitudes of \unmodeled dispersion, $0.093 z$ magnitudes of lensing dispersion \citep{holz05}, and 300 km/s peculiar velocity uncertainty. All are approximated as Gaussian and independent SN to SN.
\item $\alpha$ and $\beta$ are assumed to be constant, with values 0.13 and 3.0, respectively. $M_B$ is set to $-19.1$ and $\Omega_m$ is set to 0.3 (flat $\Lambda$CDM model).
\item For the host-mass relation, we always take $\delta(0)$ to be 0.08 magnitudes, and select $\delta(\infty)/\delta(0)$ uniformly from the range 0 to 1.
\item We assume 3\% of the SNe are outliers, and draw their observed distribution centered around the normal Ia $m_B$ for that redshift, and around zero in $x_1$ and $c$. The spread is 1, 2, and 0.5 in $m_B$, $x_1$, and $c$ respectively, and we assume these distributions are Gaussian and uncorrelated.
\item Each sample has zeropoint uncertainties of size 0.01, 0.01, 0.01, and 0.02 (highest-redshift sample) magnitudes. The uncertainties are taken to be independent sample to sample.
\item The datasets have selection effects in $m_B$, with width 0.2 magnitudes. The selection cuts are chosen for 50\% completeness at redshifts 0.08, 0.25, 0.6, and 1.45. Note that this selects from the population $x_1$ and $c$ distributions in a redshift-dependent way. (We randomly draw from the population distributions and pass them through the simulated selection effects until the required number of SNe are generated.)
\item We assume the redshift distribution of SNe scales linearly with redshift (starting from the minimum redshift of each sample). This is quantitatively incorrect (the real scaling will depend on the cosmological volume, cosmological time dilation, and the SN rate), but does produce SN samples where many SNe are removed by the magnitude cuts, allowing a good test of our selection effect modeling.
\end{itemize}

\begin{figure}
\begin{center}
\includegraphics[width = 0.45\textwidth]{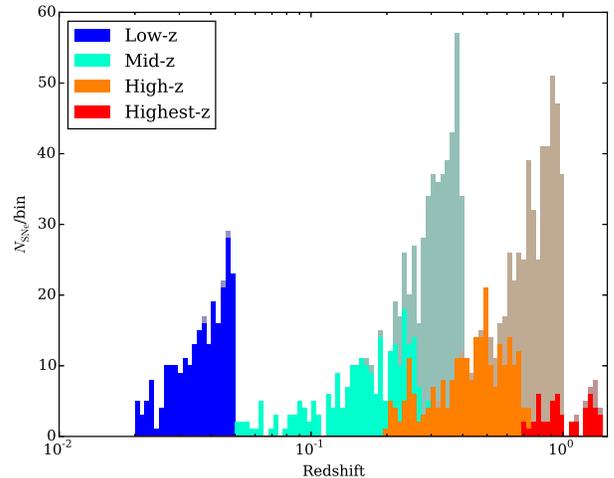}
\end{center}

\caption{Typical simulated redshift distribution for the four simulated samples in each dataset. The gray-tinted histograms show the population before selection effects. The middle two samples are essentially magnitude-limited.}
\label{fig:simdata}
\end{figure}

\begin{figure}
\begin{center}
\includegraphics[width = 0.45\textwidth]{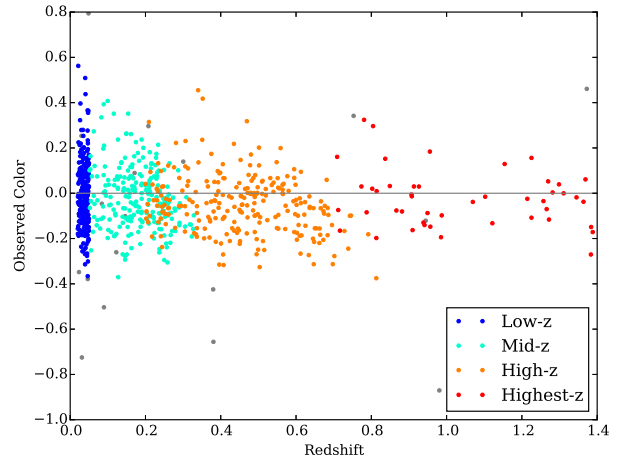}
\end{center}

\caption{Typical simulated observed color distribution for the four simulated samples in each dataset. The gray points are outliers.}
\label{fig:simcolor}
\end{figure}

\begin{figure}
\begin{center}
\includegraphics[width = 0.45\textwidth]{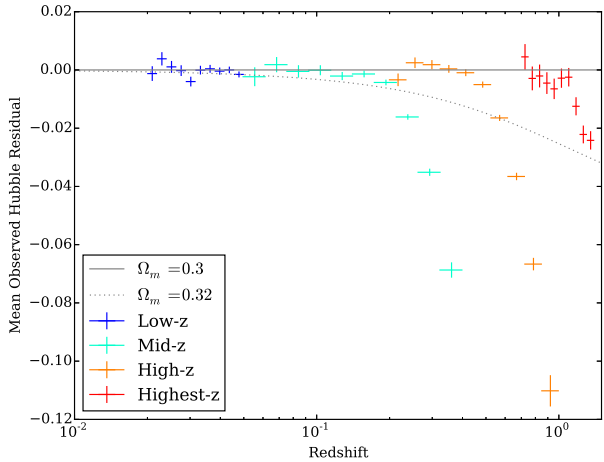}
\end{center}

\caption{Mean simulated-data Hubble-diagram residual from $\Omega_m = 0.30$ (in the sense of the numerator of Equation~\ref{eq:chi2fit}) for a large number of non-outlier points, plotted against redshift. The effect of the magnitude limits is clearly visible, and we overplot $\Omega_m=0.32$--the best-fit ignoring these selection effects.}
\label{fig:simHR}
\end{figure}

The redshift distribution of the four simulated datasets are shown in Figure~\ref{fig:simdata} before selection effects (gray-tinted histograms) and as observed. We show the observed color distributions in Figure~\ref{fig:simcolor}. This figure shows outliers in gray (knowledge that the statistical framework does not have access to), and the general trend towards bluer SNe at higher redshift due to selection effects. Figure~\ref{fig:simHR} shows the Hubble diagram residuals from the input cosmology with the best-fit ignoring selection effects ($\Omega_m = 0.32$) also shown.

After generating \nsimdata complete sets (with all four SN samples), we supply the generated SALT2 result files to the framework. Table \ref{tab:simsummary} summarizes the key results. As expected, $\alpha$, $\beta$ show bias away from zero, but any bias on $\Omega_m$ is small.

\begin{deluxetable}{ccc}
\tabletypesize{\scriptsize}
\tablecaption{Summary for primary simulated data runs.}
\tablewidth{0pt}
\tablehead{
\colhead{Parameter} & \colhead{Input Value} & \colhead{Fitted Value} \\
\colhead{} & \colhead{} & \colhead{$\pm$ Uncertainty in Mean}
}
\startdata
$\alpha$ & 0.130 & 0.143 $\pm$ 0.004 \\
$\beta$ & 3.000 &3.076 $\pm$ 0.016 \\
$M_B$ & $-19.100$ &$-$19.117 $\pm$ 0.003 \\
$\Omega_m$ & 0.300 & 0.298 $\pm$ 0.005
\enddata

\tablecomments{Average over \nsimdata simulated datasets. These results show an expected bias towards larger values of $\alpha$ and $\beta$ (discussed in Section~\ref{sec:simulateddata}), but $\Omega_m$ shows no significant bias. Other simulated data fits are also described in Section~\ref{sec:simulateddata}.}
\label{tab:simsummary}
\end{deluxetable}

We also run some simulated datasets fitting for both $\Omega_m$ and the Dark Energy equation of state parameter $w$ (assumed to be constant in redshift).\footnote{Arguably, the correct priors to use for this model are flat priors on kinematic cosmological quantities like $q_0$ and $j_0$; these priors would better preserve the Gaussian SN likelihood. The cosmological results are similar with flat priors on $\Omega_m$ and $w$, so we use flat priors on these parameters for simplicity. We constrain $\Omega_m$ to be between 0 and 1, and $w$ to be between $-2$ and 0.} We see no evidence of bias on $w$, and mild evidence of bias on the mean $\Omega_m$: $0.274 \pm 0.015$.

\section{Real Data Demonstration}\label{sec:realdata}

In this penultimate section, we demonstrate our framework on real data, namely, the Union2.1 compilation \citep{suzuki12}. This compilation is a useful dataset for demonstrating the impact of the more-sophisticated analysis, as Union2.1 provides light-curve fits for outliers \citep[the newer Joint Light-Curve Analysis,][did not publish these SNe]{betoule14}.
Our cosmological fits include $\Omega_m$ and assume a flat universe. This fit is qualitatively similar to the assumption of a constant equation-of-state parameter $w$ including a CMB or BAO constraint, in that both fits probe the deceleration parameter $q_0$. As fitting $\Omega_m$ only requires SN data, it is a cleaner analysis for our purposes here. 

\begin{figure}
\begin{center}
\includegraphics[width = 0.5\textwidth]{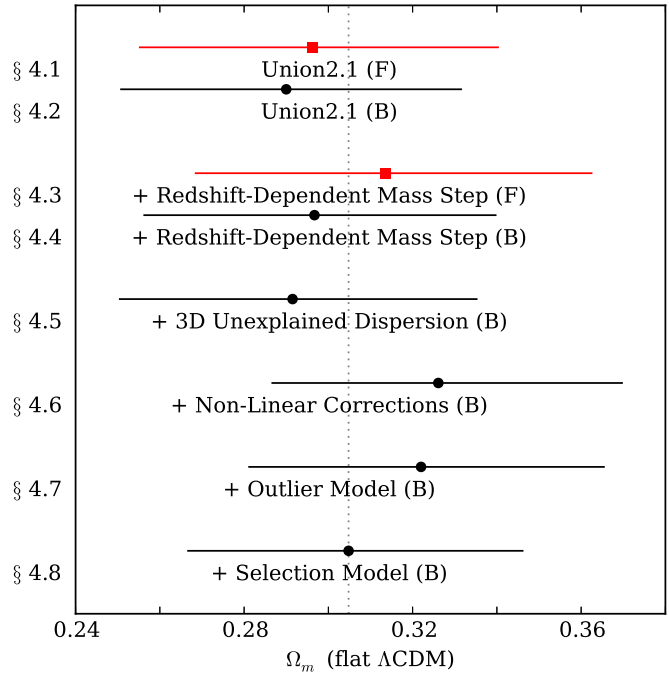}
\end{center}

\caption{Cosmological fit for each analysis. The frequentist confidence intervals show the best-fit (red squares) and the $\Delta \chi^2~=~1$ boundaries (red lines). The Bayesian credible intervals show the median of the posterior (black circles) and the 15.9 and 84.1 percentiles (black line ranges). The left margin gives the section number in the text in which each variant is discussed.}
\label{fig:analysissteps}
\end{figure}

In order to identify the effect of each feature of the analysis on the inferred results, we incrementally transition from the original Union2.1 frequentist framework to the analysis proposed in this work. The results of each step are shown in Figure~\ref{fig:analysissteps}. We conducted this part of the analysis blinded, using real data only after the code was validated on simulated data. The initial version of UNITY required the \unmodeled dispersion in $m_B$ to be fixed, which the improved selection effect model now presented in Appendix~\ref{sec:detailsselectioneffects} does not require. With the improvements in place, after a second round of blinding-unblinding, we found only a small change in $\Omega_m$ between the two versions, \Omegamchangeafterunblinding.

\subsection{Frequentist Union2.1 Analysis}
First, we show the results of a frequentist calculation, based on the same assumptions as \citet{suzuki12} and using its 580 SNe (top line in Figure~\ref{fig:analysissteps}) with SALT2 light-curve fits.\footnote{Union2.1 uses the SALT2-1 version of SALT.} All systematic uncertainties from the covariance matrix are included. To reproduce the \citet{suzuki12} results, we include only a redshift-independent host-mass \coeff.\footnote{As the \citet{suzuki12} cosmology fits including systematic uncertainties fixed $\alpha$ and $\beta$ for computational efficiency, our new results are very slightly different: 0.001 in the $\Omega_m$ confidence interval.} As a cross-check, we also try a hybrid frequentist/Bayesian model, in which the $\chi^2$ from Equation~\ref{eq:chi2fit} is converted into a likelihood as $e^{-\chi^2/2}$ and then $M_B$, $\alpha$, $\beta$, and $\delta$ are marginalized over \citep[this is similar to the method used in][]{knop03}. We obtain essentially the same results as a purely frequentist fit.

\subsection{Bayesian Model with Same Data}

For the next step, we keep all data the same, but transition to a Bayesian model for the data (the credible intervals are shown as the second line in Figure~\ref{fig:analysissteps}). This model includes the SN population terms (described in Section~\ref{sec:indpriors} and necessary for a Bayesian analysis) and the Union2.1 systematic uncertainties, but does not include our proposed treatment of selection effects, outliers, multi-dimensional \unmodeled dispersion, or the redshift-dependent host-mass \coeff. Other than the type of inference and these differences, this fit is identical to the first fit. The error bars shrink by \bayesianmodelsamedatashrink, but the central value changes very little. As the data are the same for this fit and the last one, the gain in statistical power comes from the ability of a Bayesian hierarchical model to make better use of heterogeneous information.\footnote{A Bayesian hierarchical model and a frequentist line fit will give exactly the same slope and intercept if: all uncertainties are known and Gaussian, every measurement has the same uncertainties (homoscedasticity), the Bayesian hyperpriors are Gaussian, flat priors are taken on all other parameters, and the slope and intercept are evaluated in the Bayesian model at the hyperparameter posterior maximum. These methods will give different results if the measurement uncertainties are heteroscedastic. For example, if there is a dataset with shape and color measurements but no SN has both measurements, Equation~\ref{eq:chi2fit} will have no constraining power, but a Bayesian model will. The difference between the Bayesian and frequentist Union2.1 fits is thus a measure of how inhomogeneous the uncertainties are within a given redshift range for each SN sample.}

\subsection{Frequentist Redshift-Dependent Host-Mass Standardization}

Next, we return to the frequentist framework, but include the redshift-dependent mass \coeff as described in Section~\ref{sec:hostgalaxy} (third line in Figure~\ref{fig:analysissteps}). We remove the \citet{suzuki12} covariance matrix term that corresponds to systematic uncertainty on the mass \coeff. The best fit shifts to a higher $\Omega_m$ (brighter standardized magnitudes on average at high redshift) because the high-host-mass half of the high-redshift SNe has less \coeff to fainter magnitudes. This step can be conducted through frequentist or Bayesian inference, and we show both analyses.

\subsection{Bayesian Model with Redshift-Varying Host-Mass Standardization}

In this next fit, we continue with the same data and model as the last subsection (redshift-dependent host-mass \coeff), but again transition to Bayesian inference, including the population terms (fourth line of Figure~\ref{fig:analysissteps}). The results are similar to the frequentist results, but the Bayesian fit is more agnostic about the value of $\delta(\infty)$ (essentially unconstrained), so the fit shifts less than the frequentist one to higher $\Omega_m$.

\subsection{Bayesian Model with \Unmodeled Dispersion}

Remaining in the Bayesian framework, our next addition is the multi-dimensional \unmodeled dispersion. We first remove the existing Union2.1 \unmodeled dispersion (which is only in $m_B$). By effectively increasing the uncertainties on the color, we increase the color-\coeff coefficient $\beta$. Thus the color \coeff now moves bluer (bluer due to selection effects) high-redshift SNe fainter, decreasing the fitted $\Omega_m$ (the fourth line from the bottom in Figure~\ref{fig:analysissteps}).

\subsection{Bayesian Model with Nonlinear \capitalcoeffs}

We now take advantage of our explicit $\xonetrue$ and $\ctrue$ values to include nonlinear color \coeffs\ parameterized by $d\alpha$ and $d\beta$. We remove the Union2.1 covariance terms that describe color-\coeff systematics (between the multivariate \unmodeled dispersion and the nonlinear \coeffs, these systematic uncertainties are likely much lower). The color \coeff is now strongly nonlinear (discussed further in Section~\ref{sec:otherparams}), with redder SNe requiring a larger coefficient than bluer SNe. This moves the fitted $\Omega_m$ (third line from the bottom in Figure~\ref{fig:analysissteps}) back in the opposite direction from the previous step.

\subsection{Bayesian Model with Outliers} \label{sec:modelwithoutl}

Next, we include in the fit all twelve outlier SNe removed by the Union sigma clipping in Union2.1 (a new total of 592). Instead of excluding these, we add our mixture model for handling outlier SNe. We also remove the Union2.1 systematic uncertainties on outlier rejection. The results are quite similar to the previous step, indicating that the Union sigma-clipping worked well with the 2\% contamination that was present (second line from the bottom in Figure~\ref{fig:analysissteps}).

\subsection{Bayesian Model with Selection Effects}

\begin{deluxetable*}{ccc}[h]
\tabletypesize{\scriptsize}
\tablecaption{Sample selection effects.}
\tablewidth{0pt}
\tablehead{
\colhead{Sample} & \colhead{Mag Limit} & \colhead{$m_B$ Limit ($k$-corrected at $z$)}
}
\startdata
Nearby SNe & $R$ 18.5 $\pm$ 0.5 & 18.6 $z=0.04$ \\
Cal\`{a}n/Tololo & $R$ 19.0 $\pm$ 0.5& 19.3 $z=0.1$ \\
SCP Nearby & $R$ 19.0 $\pm$ 0.5& 19.3 $z=0.1$ \\
\hline
SDSS & $r$ 22.1 $\pm$ 0.5 (AB) & 22.6 $z=0.4$\\
SNLS & $i$ 24.3 $\pm$ 0.5 (AB) & 25 $z=1$\\
Other mid-redshift & $R$ 24 $\pm$ 0.5 & 24.5 $z=0.8$ \\
\hline
High-redshift ground & $I$ 23.8 $\pm$ 0.5 & 25.0 $z=1.0$\\
HST ACS & $z$ 25.0 $\pm$ 0.5 (Vega) & 26.1 $z=1.5$
\enddata

\tablecomments{Approximate magnitude limits for each sample. We use a SALT2 $c=0$ SN template to convert each magnitude limit to the limit in rest-frame $B$-band magnitude at the specified redshift.}
\label{tab:selection}
\end{deluxetable*}

Finally, we model selection effects. For simplicity, we approximate all selection as occurring in rest-frame $B$-band ($a_j^{\mathrm{cut}} = 0$, $b_j^{\mathrm{cut}} = 0$), and leave a more detailed analysis to future work.\footnote{As the present paper is primarily focused on methods, a very detailed treatment of the data is beyond its scope.} For many of the most important samples in Union2.1, this is not a bad approximation. For example, the Sloan Digital Sky Survey SNe \citep{holtzman08} were selected in $g$, $r$, and $i$-band \citep{dilday08}. At redshift $\sim 0.4$ (the distant end of the survey), $r$-band corresponds to rest-frame $B$-band. The Supernova Legacy Survey SNe were selected in $i$-band \citep{perrett12}, which matches rest-frame $B$-band for the highest-redshift SNe with small distance uncertainties ($z \sim 0.7$).

For some surveys, rest-frame $B$-band selection is a poor approximation. Many of the nearby SNe were selected from unfiltered surveys (approximately rest-frame $R$-band), but these mostly galaxy-targeted surveys have generally weak selection effects in magnitude (discussed below). Some distant SN surveys \citep{tonry03, riess07, amanullah10, suzuki12, rubin13} had selections that were different (bluer) than rest-frame $B$-band. At least for the HST-discovered SNe, selection effects are small for most of the redshift range (also discussed more below).

We estimate the selection effects for each sample as follows (summarized in Table~\ref{tab:selection}). Nearby SNe are generally limited by spectroscopic followup, for example the $\sim 18.5$ magnitude limit for CfA discussed in \citet{hicken09} (although this is unfiltered, we approximate it as $R$-band). We take this limit as typical. The Cal\`{a}n/Tololo survey \citep{hamuy96a, hamuy96b} and the SCP nearby search \citep{kowalski08} extend out to higher redshift; the limiting magnitude in this case is $R \sim 19$ \citep{hamuy99}. We take the magnitude limit for SDSS from \citet{dilday08} and the limit for SNLS from \citet{perrett12}. Together, these samples make up most of the mid-redshift weight. For the other mid-redshift samples, we take a limit of $R$ = 24, judged from the approximate rolloff of the SN population in redshift. For the high-redshift ground-discovered samples, we assume the surveys were 50\% complete at $z=1$; this gives a limit of 23.8 in $I$-band. Finally, for the HST-discovered SNe, we take a limit from \citet{barbary12} of $z$-band $\sim 25$ (Vega). In all cases, we take the width of the selection to be $\pm 0.5$ magnitudes (i.e., a SN 0.5 magnitudes brighter than the mean cut has an 84\% chance of being selected; a SN 0.5 magnitudes fainter has a 16\% chance). As a cross-check, we coherently shift each estimated magnitude limit fainter by 0.5 magnitudes, representing an extreme limit of how inaccurate our estimations are likely to be. The $\Omega_m$ credible interval shifts by \OmegamShiftMagLim.

We remove the Union2.1 covariance matrix terms for Malmquist bias before computing the fit shown in the bottom line of Figure~\ref{fig:analysissteps}. The $\Omega_m$ credible interval shifts to lower $\Omega_m$, as the distant SNe with significant selection effects are standardized fainter. The central value of this final fit closely matches the original Union2.1 result, but we see that our new, smaller (by \stepsevensmaller compared to the Union2.1 analysis) credible interval shows an increase in statistical power. It is worth reiterating that the list of improvements to make was established before the results were known; we did not set out to simply achieve a similar result to Union2.1. The scatter of the intermediate results generally validates the size of the Union2.1 systematics estimates for these effects that we can now properly include in the model.

\subsection{Other Parameters}\label{sec:otherparams}

We now present the results for important nuisance parameters and their relation to $\Omega_m$ in the form of 1D and 2D credible regions. In both cases, our credible regions are derived by using Kernel Density Estimation with a Gaussian kernel on the MCMC samples,\footnote{We draw $\sim$3,000 samples from sixteen chains. The \citet{gelmanrubin92} $R$ statistics are $\lesssim 1.01$, indicating good convergence.} then solving for the contour level that encloses 68.3\% (inner shaded regions) and 95.4\% (outer shaded regions) of the posterior.\footnote{Kernel Density Estimation gives biased results when the posterior samples lie against a parameter boundary, as the density will be smoothed including regions with zero density. For all our variables, we reflect the nearby samples through the minimum and maximum samples, creating a virtual dataset on the other side of such boundaries.}

\begin{figure*}
\begin{center}
\includegraphics[width = 0.9\textwidth]{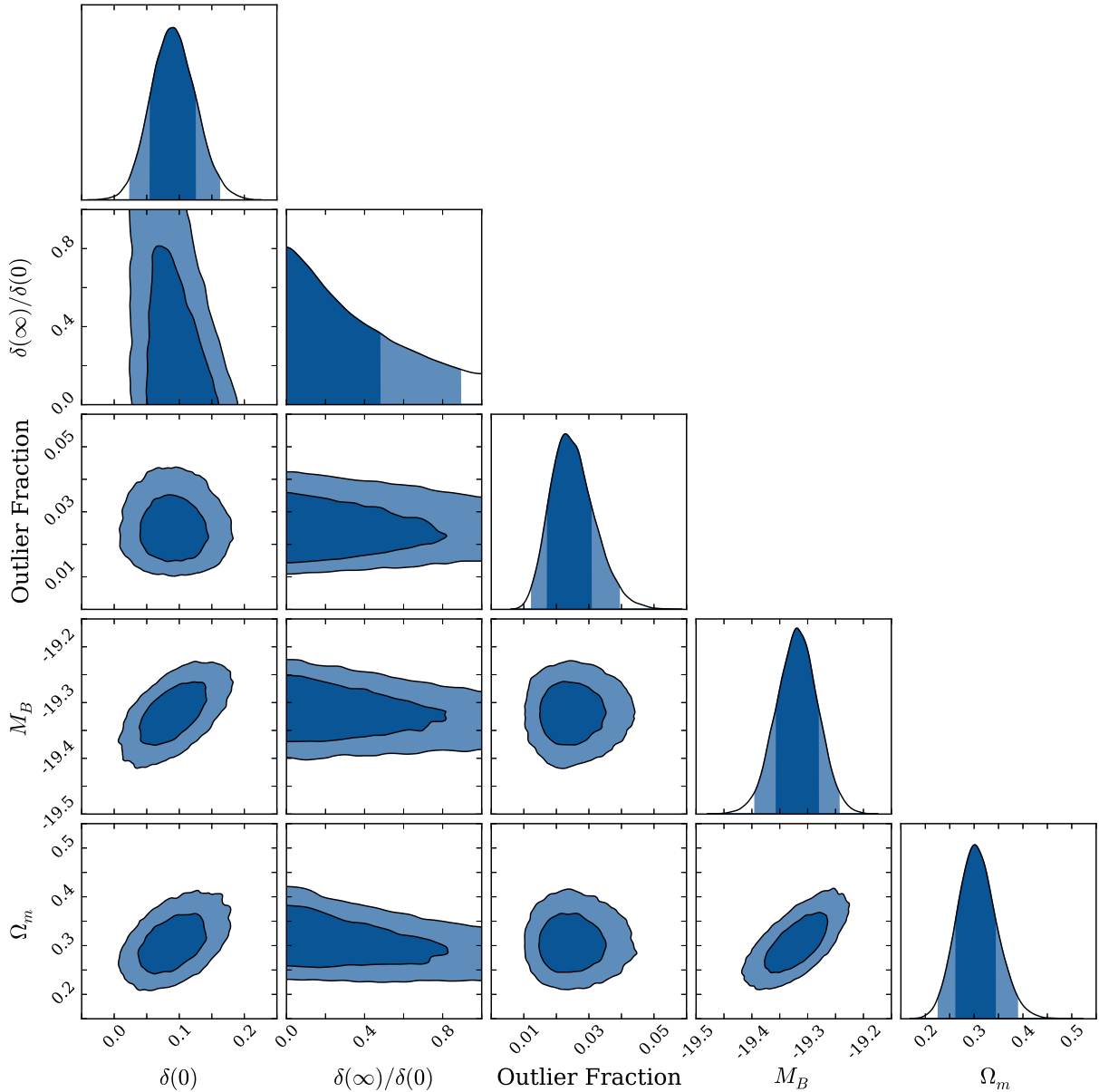}
\end{center}

\caption{Credible intervals and regions of the mass-\coeff coefficient $\delta(0)$, its variation with redshift, $\delta(\infty)/\delta(0)$, the estimated outlier fraction, $M_B$, and $\Omega_m$. The dark shading is the 68.3\% credible region; the lighter shading is 95.4\%. The mass-\coeff coefficients have a significant correlation with cosmological parameters, but the estimated outlier fraction does not.}
\label{fig:mass}
\end{figure*}

Figure~\ref{fig:mass} shows the significant degeneracy between both host-mass \coeff parameters and $\Omega_m$, illustrating that neither one should be neglected. The mean value of the estimated fraction of outliers is similar to the 12/592 found by the Union sigma clipping. This parameter has no significant degeneracy with any parameter for the spectroscopically confirmed datasets in Union2.1, confirming that outlier rejection is not a significant concern at this high level of purity. As an additional cross-check, instead of assuming the outlier distribution is centered on the SN Ia distribution, we fit for it. Including six parameters in the model for the mean and dispersion in \{$m_B$, $x_1$, $c$\} (taken to be uncorrelated, and assuming a constant distribution with redshift) leaves the error bar unchanged but shifts the credible region by \outlierdistOmshift in $\Omega_m$. If future versions of the UNITY framework are run on samples with larger contamination, a more flexible outlier parameterization can be matched with the increased number of outliers in the fit \citep[e.g.,][]{hlozek12}.

\begin{figure*}
\begin{center}
\includegraphics[width = 0.9\textwidth]{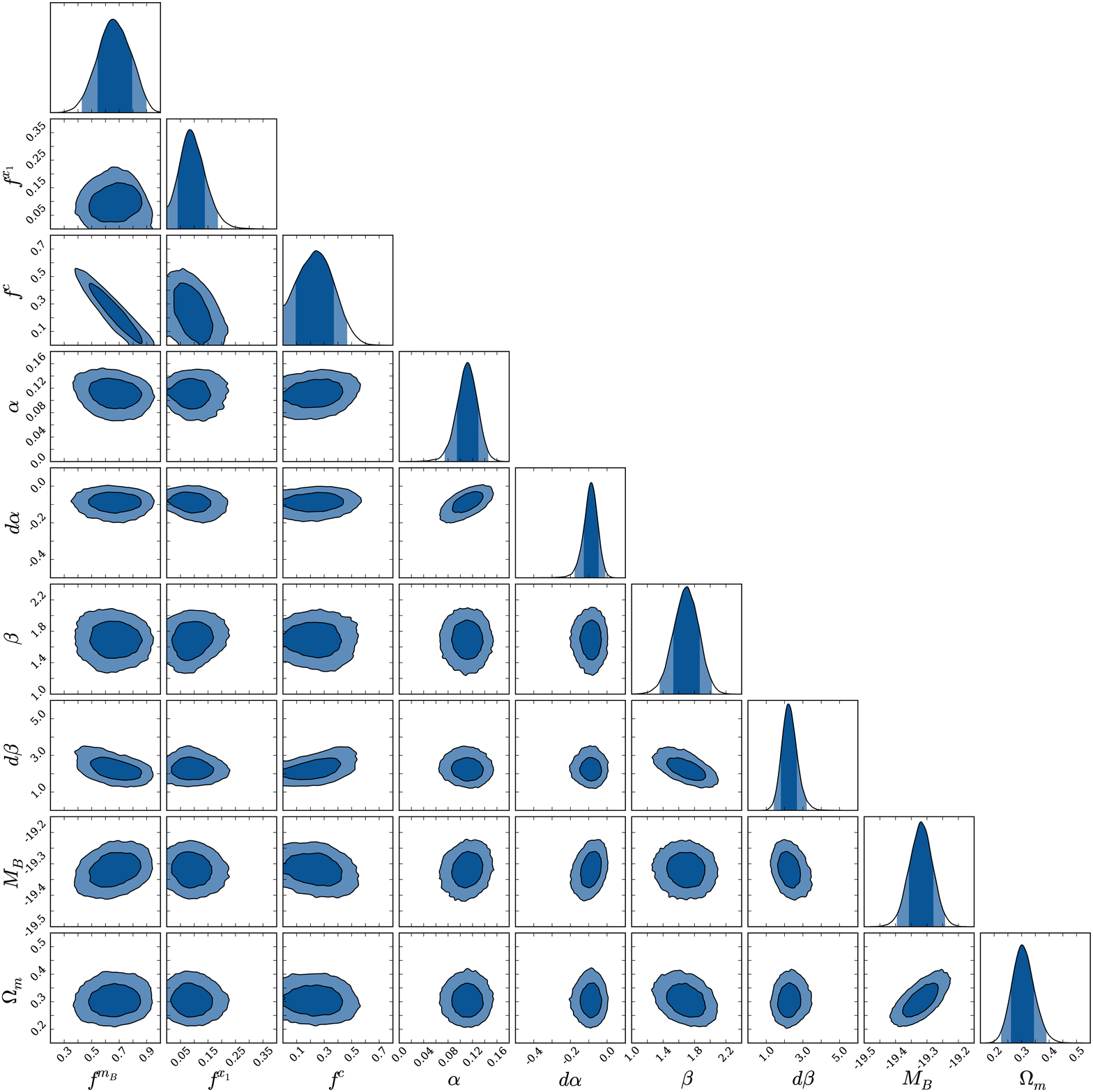}
\end{center}

\caption{Credible intervals and regions of the \unmodeled variance in \{$m_B$,~$x_1$,~$c$\}, the \coeff coefficients, and the cosmological parameters. The dark shading is the 68.3\% credible region; the lighter shading is 95.4\%.}
\label{fig:intalphabeta}
\end{figure*}

Figure~\ref{fig:intalphabeta} shows the degeneracy between the fraction of the \unmodeled variance in $x_1$ and $\alpha$, and similarly $c$/$\beta$. Including these \unmodeled dispersion parameters increases the uncertainty and the mean value for $\alpha$ and $\beta$. (Here, $\alpha$ and $\beta$ represent the mean \coeff coefficient.) The model prefers most of the \unmodeled dispersion in  $m_B$ and $c$, rather than $x_1$. We also see statistical evidence of non-zero $d\alpha$ and $d\beta$ (recall that $d\alpha$ is the $x_1$ \coeff coefficient for broad-light-curve SNe minus the coefficient for narrow SNe; $d\beta$ is similarly defined for red minus blue SNe): $d\alpha$ is \dalphaconstraint and $d\beta$ is \dbetaconstraint.\footnote{This extreme difference in the color-standardization relation with color likely implies that SALT2 should be retrained with a color-dependent color law.} Both have a correlation with $\Omega_m$, although the correlation with $d\beta$ is larger. We note, at least for this compilation of SNe and SALT2-1, that blue SNe still have a non-zero $\beta$: \betablueconstraint, and red SNe have a $\beta$ of \betaredconstraint. This value is significantly less than 4.1 (the value expected if all reddening were due to the mean extinction law of the Milky Way diffuse interstellar medium).\footnote{Although we divide the broken-linear relation at $c=0$ in our standard analysis, letting the division be a fit parameter yields a division at \CZeroConstraint; the uncertainties on $\beta^B$ significantly increase.}

\newcommand{\sigxoneintsq}{\sigma_{x_1 \mathrm{int}}^2}
\newcommand{\sigcintsq}{\sigma_{c \mathrm{int}}^2}

\section{Conclusions} \label{sec:conclusions}

In this work, we propose UNITY, a unified Bayesian model for handling outliers, selection effects, shape and color \coeffs, \unmodeled dispersion, and heterogeneous observations. We demonstrate the method with the Union2.1 SN compilation, and show that our method has \stepsevensmaller smaller uncertainties, but results that are consistent with the Union2.1 analysis. The advantages of UNITY will likely be even larger in upcoming datasets, as scarce followup resources will introduce heterogeneity, and enlarged samples may reduce the cosmological-parameter statistical uncertainties below the size of our improvements.

There are several future directions for this research. We could allow the \unmodeled dispersion covariance matrix to vary with other parameters ($x_1$, $c$, redshift, or rest-frame wavelength range). We could decompose the \unmodeled dispersion into a sum of Gaussians as in the model of \citet{rigault13}. We could use Gaussian process regression \citep{gpml06} to handle the redshift-dependent priors on $x_1$ and $c$. Our model also lets us include more than one light-curve fit for each SN, with the covariance matrix between different light-curve fitters parameterized. We could even make use of the hierarchical model to constrain ``exotic possibilities,'' as sketched in Section~\ref{sec:desiredprops}. All of these are straightforward modifications of what we present here, but our proposed model is already superior to current SN cosmological analysis frameworks. We believe these concepts will also be useful for other applications.

\acknowledgments

We thank Alex Kim, Marisa March, Masao Sako, Rachel Wolf, and the Referee for their feedback on this manuscript. This work was supported in part by the Director, Office of Science, Office of High Energy Physics, of the U.S. Department of Energy under contract No. DE-AC02-05CH11231. This research used resources of the National Energy Research Scientific Computing Center, a DOE Office of Science User Facility supported by the Office of Science of the U.S. Department of Energy under Contract No. DE-AC02-05CH11231.

\appendix

\section{Details of the Full Framework} \label{sec:frameworkdetail}

\newcommand{\fracdet}{P(\mathrm{detect})\xspace}
\newcommand{\fracdeti}{P(\mathrm{detect}|z_i)\xspace}
\newcommand{\fracdetobs}{P(\mathrm{detect}|\{\mBobsi, \xoneobsi, \cobsi\})\xspace}
\newcommand{\Pobsparams}{P(\{\mBobsi, \xoneobsi, \cobsi\}|\mathrm{params})\xspace}

Our model is a two-component mixture model, where the outlier component is a Gaussian in $m_B$, $x_1$, and $c$, and the normal SNe Ia component is a Gaussian in $m_B$, $x_1$, and $c$, truncated by selection effects. The full likelihood for a single SN is given by the sum of the core distribution
\newcommand{\mcut}{m^{\mathrm{cut}}}

\begin{equation}
\frac{\mathcal{N} \left( \left[
\begin{array}{l}
\mBobsi \\
\xoneobsi \\
\cobsi
\end{array}
\right]
,
\left[ \begin{array}{l}
\mBtruei + \Delta m_B\\ 
\xonetruei +  \Delta x_1\\
\ctruei + \Delta c
\end{array}
\right], \  C^{\mathrm{ext}}(z_i) + C^{\mathrm{obs}}_i + C^{\intr}_i \right) \mult \fracdetobs}
{(\epsilon=0.01) +  \fracdet} (1 - f^{\mathrm{outl}})\\
\end{equation}

with the outlier distribution

\begin{equation}
\mathcal{N} \left( \left[
\begin{array}{l}
\mBobsi \\
\xoneobsi \\
\cobsi
\end{array}
\right]
,
\left[ \begin{array}{l}
\mBtruei + \Delta m_B\\ 
\xonetruei +  \Delta x_1\\
\ctruei + \Delta c
\end{array}
\right], \  ( \mathbbm{1}  )  \right) f^{\mathrm{outl}} \;.
\end{equation}

$\fracdetobs$ and $\fracdet$ are described in Equations~\ref{eq:fracdetobs} and \ref{eq:Psel}, respectively. $\mBtruei$ is not a free parameter; instead it is completely determined by other parameters and is given by

\begin{equation}
\mBtruei \equiv M_B -\alpha (\xonetruei) \mult \xonetruei + \beta (\ctruei) \mult \ctruei + \delta(z_i) \mult (M_{\star} > 10^{10} M_{\odot}) + \mu(z_i, \mathrm{cosmo})\;.
\end{equation}

$\Delta$\{$m_B$, $x_1$, $c$\} are the contributions from the systematic uncertainty terms, e.g., 
\begin{equation}
\Delta m_B \equiv \sum_l \frac{\partial \mBobsi}{\partial \Delta \mathrm{sys}_l} \Delta \mathrm{sys}_l \;.
\end{equation}

The broken-linear slopes are defined as

\begin{equation}
\alpha (\xonetruei) \equiv
\begin{cases}
\alpha^S,& \xonetruei < 0 \\
\alpha^L,& \xonetruei > 0 \\
\end{cases}
\end{equation}

\begin{equation}
\beta (\ctruei) \equiv
\begin{cases}
\beta^B,& \ctruei < 0 \\
\beta^R,& \ctruei > 0 \\
\end{cases}
\end{equation}

and $\delta(z_i)$ is given by Equation~\ref{eq:deltaofz}.

We take the following priors on all parameters:

\begin{align}
M_B & \sim \mathcal{U}(-20, -18) & \text{Absolute Magnitude for } h = 0.7 \notag \\
\tan^{-1}{\alpha^S} & \sim \mathcal{U}(-0.2, 0.3) & \text{$x_1$ \capitalcoeff Coefficient, $\xonetruei < 0$} \notag \\
\tan^{-1}{\alpha^L} & \sim \mathcal{U}(-0.2, 0.3) & \text{$x_1$ \capitalcoeff Coefficient, $\xonetruei > 0$} \notag \\
\tan^{-1}{\beta^B} & \sim \mathcal{U}(-1.4, 1.4) & \text{$c$ \capitalcoeff Coefficient, $\ctruei < 0$} \notag \\
\tan^{-1}{\beta^R} & \sim \mathcal{U}(-1, 1.4) & \text{$c$ \capitalcoeff Coefficient, $\ctruei > 0$} \notag \\
\delta(0) & \sim \mathcal{U}(-0.2, 0.3) & \text{Host-Mass \capitalcoeff Coefficient, $z=0$} \notag \\
\delta(\infty) & \sim \mathcal{U}(0, 1) \delta(0) & \text{Host-Mass \capitalcoeff Coefficient, $z=\infty$} \notag \\
\Omega_m & \sim \mathcal{U}(0, 1) & \Omega_m \text{ (flat $\Lambda$CDM)} \notag \\
\log_{10}{\sigma^{\intr}_j} & \sim \mathcal{U}(\log_{10}{0.05}, \log_{10}{0.50}) & \text{``Sample'' (\Unmodeled) Dispersion} \notag \\
f^{m_B}, f^{x_1}, f^c & \sim \mathcal{U}(0, 1) & \text{Fraction of \Unmodeled Dispersion in $\{m_B, x_1, c\}$} \notag \\
\rho^{\intr}  & \sim \mathrm{LKJ}(1) & \text{Correlation Matrix of \Unmodeled Dispersion} \notag \\
\xonetruei & \sim \mathcal{N}(x_{1jk}^*, (R_j^{x_1})^2) & \text{Modeled Latent $x_1$} \notag \\
\ctruei & \sim \mathrm{SkewNormal}(\mathrm{mean} = c_{jk}^*, \mathrm{variance} = (R_j^c)^2, \alpha = \alpha_{jk}^{\mathrm{S-N}}) & \text{Modeled Latent $c$}\notag \\
\delta^{\mathrm{S-N}}_{jk} & \sim \mathcal{U}(-0.995, 0.995) & \text{Related to SkewNormal Shape Parameter} \notag \\
\log_{10}{R_j^{x_1}} & \sim \mathcal{U}(-0.5, 0.5) & \text{Dispersion of Latent $x_1$} \notag \\
\log_{10}{R_j^c} & \sim \mathcal{U}(-1.5, -0.5) & \text{Dispersion of Latent $c$}\notag \\
x_{1jk}^* & \sim \mathrm{Cauchy}(0, 1) \notag & \text{Mean of Latent $x_1$} \\
c_{jk}^* & \sim \mathrm{Cauchy}(0, 0.3) \notag& \text{Mean of Latent $c$} \\
\log{f^{\mathrm{outl}}} & \sim \mathcal{N}(-3, 0.5^2) & \text{Fraction of Outliers} \notag 
\end{align}

\section{Details of the Selection-Effect Model} \label{sec:detailsselectioneffects}

In this section, we detail the approach we take for modeling selection effects in UNITY. We start with a general likelihood containing missing observations in the case when both the number of observed objects ($\nobs$) and the number of missed objects ($\nmiss$) are exactly known:

\begin{equation} \label{eq:fracdetect}
\left( \begin{array}{c}
\nobs+\nmiss \\
\nobs \end{array} \right) \left[ \prod_{i=1}^{\nobs} \Pobsparams \fracdetobs \right] \left[ \prod_{i=1}^{\nmiss} (1 - \fracdet) \right]\;.
\end{equation}

As described in Section~\ref{sec:selection}, we assume that the efficiency smoothly varies with $\{\mBobsi, \xoneobsi, \cobsi\}$:
\begin{equation} \label{eq:fracdetobs}
\fracdetobs = \Phi \left( \frac{ \mcut - (\mBobsi + a_j^{\mathrm{cut}}  \xoneobsi +  b_j^{\mathrm{cut}} \cobsi) } {\sigma_{\mathrm{cut}}} \right) \;,
\end{equation}
where $\Phi$ is the Gaussian CDF. Of course, we do not know any of the parameters for each of the missing observations. $\fracdet$ is thus marginalized over the entire distribution when it is not referencing a specific observation.

We now make the counterintuitive approximation that the redshift of each missed SN is exactly equal to the redshift of a detected SN. This approximation is accurate because the SN samples have, on average, enough SNe that the redshift distribution is resonably sampled. We now refactor Equation~\ref{eq:fracdetect} to consider each detected SN:

\begin{equation}
\prod_{i=1}^{\nobs}  \left( \begin{array}{c}
1+\nmiss_i \\
1
\end{array} \right)  \Pobsparams \fracdetobs (1 - \fracdeti)^{\nmiss_i}\;.
\end{equation}

The combinatoric factor trivially becomes $(1 + \nmiss_i)$. To minimize the number of parameters in the cosmological fit, we now seek to marginalize over $(1 + \nmiss_i)$ from 1 to $\infty$. It is common \citep[e.g., ][]{gelman2013bayesian, kelly07} to take a flat prior on $\log{(\nmiss_i)}$; for reasons that will become clear in a moment, we force this prior to zero faster than this by multiplying this $(1 + \nmiss_i)^{-1}$ prior by a weak exponential decay
\begin{equation}
P(1 + \nmiss_i) = \frac{\exp{(-\epsilon (\nmiss_i + 1))}}{1 + \nmiss_i}\;,
\end{equation}
where $\epsilon$ is a small positive number. Thus, the marginalization becomes:

\begin{equation}
\prod_{i=1}^{\nobs} \sum_{(1 + \nmiss_i) = 1}^{\infty} \Pobsparams \fracdetobs (1 - \fracdeti)^{\nmiss_i} \exp{(-\epsilon (\nmiss_i + 1))}\;
\end{equation}
which is equal to the geometric series
\begin{equation}
\prod_{i=1}^{\nobs} \sum_{(1 + \nmiss_i) = 1}^{\infty} \Pobsparams \fracdetobs ((1 - \fracdeti) \exp{(-\epsilon}))^{\nmiss_i} \exp{(-\epsilon})\;
\end{equation}
which sums to 
\begin{align}
& \prod_{i=1}^{\nobs} \frac{\Pobsparams \fracdetobs}{\exp{(\epsilon}) - (1 - \fracdeti)} \nonumber \\ 
 \approx & \prod_{i=1}^{\nobs} \frac{\Pobsparams \fracdetobs}{\epsilon + \fracdeti} \;
\end{align}
This expression highlights the benefit of this prior. If a flat prior on $\log{(\nmiss_i)}$ is assumed ($\epsilon = 0$), then the likelihood can be poorly behaved in regions of parameter space where the efficiency is poor (even if care is taken in the log likelihood to correctly handle the asymptotes of the log of the efficiency). Running UNITY on Union2.1 with $\epsilon$ either 0 or 0.01 makes virtually no difference, but Stan has difficulty with many of the simulated datasets (which we constructed with severe Malmquist bias as a test, see Section~\ref{sec:simulateddata}) for $\epsilon = 0$, so we make $\epsilon = 0.01$ our default choice.

We now need the efficiency as a function of the parameters. We derived the following expression by analytically computing the variance, then expanding to first order in $x_1^*$ (this approximation is good to better than one percent in the variance for our distributions). For simplicity, we use the same expression for $c$, although the $c$ distribution is skew-normal instead of normal. (Changing the assumed intrinsic color distribution to normal, rather than skew-normal, changes $\Omega_m$ by \Omegamnoskew, so this approximation is not a significant concern.)

\begin{align}
V^{m_B} & = C^{m_B}_i + \sigma^2_{\mathrm{cut}} + (\delta(z_i)/2)^2  \\
               & + ((\alpha -a_j^{\mathrm{cut}}) \mult R^{x_1})^2 + (\alpha -a_j^{\mathrm{cut}}) \mult d\alpha \mult {x_1^*} \mult R^{x_1} \sqrt{\frac{2}{\pi}} + (d\alpha \mult R^{x_1})^2 \frac{\pi - 2}{4 \pi} \nonumber \\
               & + ((\beta + b_j^{\mathrm{cut}}) \mult R^c)^2 + (\beta + b_j^{\mathrm{cut}}) \mult d\beta \mult c^* \mult R^c \sqrt{\frac{2}{\pi}} + (d\beta \mult R^c)^2 \frac{\pi - 2}{4 \pi} \nonumber \;.
\end{align}

The first line contains the measurement and \unmodeled dispersion in $m_B$, the dispersion of the selection cut, and the dispersion due to the host-galaxy relation. The second and third lines give the dispersion due to the $x_1-m_B$ and $c-m_B$ relations, respectively. Using the same approximations, we derived an expression for the mean of the (unstandardized) magnitude distribution

\begin{equation}
\mathrm{mean}  = M_B + \mu(z_i) + (b_j^{\mathrm{cut}}  + \beta)\mult c^* + \frac{d\beta \mult R^c}{2\pi} - (-a_j^{\mathrm{cut}} + \alpha) \mult x_1^*  - \frac{d\beta \mult R^{x_1}}{2\pi}\;;
\end{equation}
in this case, our approximations are valid to a few hundredths of a magnitude. We then compute the selection efficiency assuming that the distribution of magnitudes is Gaussian
\begin{equation} \label{eq:Psel}
\fracdeti = \Phi((m_B^{\mathrm{cut}} - \mathrm{mean})/\sqrt{V^{m_B}})\;.
\end{equation}
This approximation is not completely valid, as there is a tail to faint magnitudes of red SNe, but for the central part of the distribution (which is the part that is cut into by severe Malmquist bias), it is a good approximation.

\end{document}